\UseRawInputEncoding
\documentclass[aps,prd,onecolumn,groupedaddress,showpacs,nofootinbib,amssymb]{revtex4}
\usepackage[dvips]{graphicx,color}
\usepackage{amssymb}
\usepackage{amsmath}
\usepackage{graphicx}
\usepackage{amsfonts}

\newcommand{\be}{\begin{equation}}
\newcommand{\ee}{\end{equation}}
\newcommand{\bea}{\begin{eqnarray}}
\newcommand{\eea}{\end{eqnarray}}
\newcommand{\beaa}{\begin{eqnarray*}}
\newcommand{\eeaa}{\end{eqnarray*}}

\begin{document}

\title{Gravitational baryogenesis in DGP brane cosmology}
\author{K. Atazadeh }\email{atazadeh@azaruniv.ac.ir}
\affiliation{Department of Physics, Azarbaijan Shahid Madani University, Tabriz, 53714-161 Iran}

\begin{abstract}
We consider the imbalance of matter and antimatter by using gravitational baryogenesis mechanism in the background of DGP brane cosmology. By taking into account a flat Friedmann-Lema\^{\i}tre-Robertson-Walker (FLRW) metric in DGP brane model, we obtain that for a radiation dominated universe, $w=1/3$, the ratio of baryon number density to entropy from the gravitational baryogenesis is not zero, contrary to ordinary general relativity.
Also, we study the ratio of baryon to entropy against the observational constraints in the DGP cosmology.
\end{abstract}


\maketitle

\section{Introduction}

The DGP brane world model assumes the existence of a $5$-dimensional Minkowski space, within which ordinary $4$-dimensional Minkowski space is embedded. The model introduces an action consisting of two terms, one term is the usual Einstein–-Hilbert action, which involves only the $4$-dimensional spacetime dimensions. The other term is the equivalent of the Einstein–-Hilbert action, as extended to all $5$-dimensions. The $4$-dimensional term dominates at short distances, and the $5$-dimensional term dominates at long distances~\cite{devali}. However, there are two different approaches to embed the $4$-dimensional brane into the $5$-dimensional spacetime, the DGP model has two separate branches marked by $\epsilon$ with distinct aspects. The $\epsilon = +1$ branch can illustrate the present accelerating cosmic expansion without introducing a mysterious fluid called dark energy \cite{Deffayet}, while for the $\epsilon = -1$ branch, dark energy is needed in order to yield an accelerating expansion \cite{Lue,26}. Moreover, inflation in the DGP model displays some new characteristics. It must be noted that only in the $\epsilon = -1$ case can inflation exit spontaneously \cite{30,31,32,33,34,35}.

The gravitational baryogenesis (GB) process was proposed in  2004 by  Davoudiasl {\it et al} \cite{Davoudiasl}, in order to explain the observed imbalance of matter over antimatter in the observable universe. The cosmic microwave background observations \cite{Bennett} and also the Big Bang nucleosynthesis \cite{Burles} admite this imbalance of matter (baryons) and antimatter (antibaryons). According to Sakharov benchmark \cite{sakhar}, the GB procedure is based on the presence of a $CP$-violating interaction term that it can explain the fundamental conditions for the generation of the imbalance of matter and antimatter in the observed universe. In this paper we try to consider the GB process in the context of DGP brane cosmology, and the goal of the paper is to highlight the differences of GB process between the DGP brane cosmology and standard cosmology. However, the imbalance of matter and antimatter GB term is given by \cite{Davoudiasl}
\begin{equation}
\label{1}
\frac{1}{M_{*}^2}\int d^4x\sqrt{-g}(\partial_{\mu}R)J^{\mu}\, ,
\end{equation}
this term can be resulted from the higher-order interactions in the gravitational physics. In the equation (\ref{1}), $M_*$ is the cutoff scale of the underlying effective gravitational theory, $J^{\mu}$ stands for the baryonic matter current and finally $g$ is the determinant of the metric $g_{\mu \nu}$ and also $R$ is the Ricci scalar. In the general relativity (GR) regime for flat FLRW metric the ratio of baryon number density to entropy, $n_B/s$, for radiation dominated universe is zero, thus at the early times the GB process can not generate baryonic matter asymmetry. However, we show in that in the context of DGP brane cosmology, the ratio of baryon number density to entropy differs from zero in the radiation dominated universe. We try to find the general expression for the ratio of baryon number density to entropy in DGP brane world and we show that results can be compatible with the observational constraint $n_{B}/s<9 \times 10^{-11}$.

GB mechanism has been studied in various theories: in $f(R)$ gravity~\cite{Lambiase,Ramos}, in brane world scenarios~\cite{Koyama}, in loop quantum cosmology~\cite{Odintsov}, in $f(T)$ gravity~\cite{sari}, in Gauss-Bonnet brane world cosmology~\cite{Bento}, in Gauss-Bonnet gravity~\cite{Odintsov2}, in running vacuum models~\cite{Oikonomou}, in Ho\v{r}ava gravity \cite{Rudra} and also see Refs.\cite{1,2,3,4,5,6}.

This paper is organized as follows: In section II, we consider the modified Friedmann equations in the framework of DGP brane universe and we take the cosmological evolution of the model in the ultra hight energy limit. In section III we study in detail the cases in which the resulting ration of baryon number density/entropy can be compatible with the observational data. Conclusions are drawn in the last section.

\section{The friedmann equation in DGP brane world}

A homogeneous and isotropic universe can be described by the FLRW line element, as follows
\begin{eqnarray}
ds^{2}=-dt^2+a^{2}(t)\left(\frac{dr^{2}}{1-kr^{2}}+r^{2}d^{2}\Omega\right)\;.
\end{eqnarray}
One can obtain the Friedmann equations on the warped DGP brane as~\cite{Maeda,wu}
\begin{eqnarray}
H^{2}+\frac{k}{a^{2}}=\frac{1}{3\mu^{2}}[\rho+\rho_{0}(1+\epsilon
\chi(\rho, a))]\;,
\end{eqnarray}
where $\mu$ is a parameter standing for the strength of the
induced gravity  on the brane, $H$ is the Hubble parameter, $k$ is the spatial curvature of
the FLRW metric and $\rho$ denotes for the total energy density.
For $\epsilon=+1$, the brane tension is negative,
while for $\epsilon=-1$, it is positive. $\chi$ is given by
\begin{eqnarray}
\chi=\left[\chi_{_{0}}^2+\frac{2\eta}{\rho_{0}}\left(\rho-\mu^2\frac{\mathcal
{E}_0}{a^4}\right)\right]^{1/2}\;,
\end{eqnarray}
where \begin{eqnarray}
\chi_{_{0}}=\sqrt{1-2\eta
\frac{\mu^{2}\Lambda}{\rho_{0}}},\quad \eta=\frac{6m_{_{5}}^6}{\rho_{0}\mu^{2}}
\;\;\; (0<\eta\leq 1), \quad \rho_{0}=m_{_{\lambda}}^4+6\frac{m_{_{5}}^6}{\mu^{2}}\;,
\end{eqnarray}
where $\Lambda$ is given by
\begin{eqnarray}
\Lambda=\frac{1}{2}(^{(5)}\Lambda+\frac{1}{6}\kappa_{_{5}}^6\lambda^{2})\;.
\end{eqnarray}
where $^{(5)}\Lambda$ is the $5$-dimensional cosmological constant in the bulk,
$\lambda$ is the brane tension, $\kappa_{_{5}}$ stands for the $5$-dimensional Newton constant and $\mathcal {E}_0$ is a constant
related to Weyl radiation. In order to more simplicity, we will put $\Lambda=0$, thus the equation (2) can be written as
\begin{eqnarray}
H^{2}+\frac{k}{a^{2}}=\frac{1}{3\mu^{2}}\left(\rho+\rho_{0}+\epsilon\rho_{0}\sqrt{1+\frac{2\eta\rho}{\rho_{0}}}\right).
\end{eqnarray}
To discuss the GB mechanism in the very early era of the universe in where the total energy
density is very high, thus we will consider only the ultra high energy limit, $\rho$$\gg$$\rho_0$. So, for a flat FLRW universe ($k=0$), the Friedmann equation is given by
\begin{eqnarray}\label{2}
H^{2}=\frac{1}{3\mu^{2}}(\rho+\epsilon\sqrt{2\rho\rho_{0}}).
\end{eqnarray}
This equation is modification of $4$-dimensional gravity with minor corrections, where $\mu$ must have an energy scale as the Planck scale in
the DGP model.

The conservation equation for the universe with the perfect fluid, is given by
\begin{eqnarray}\label{3}
\dot{\rho}=-3H(1+w)\rho,
\end{eqnarray}
where $w$ is the equation of state parameter.

Differentiating equation (\ref{2}) with respect to cosmic time gives
\begin{eqnarray}\label{5}
\dot{H}=-\frac{1}{2\mu^{2}}(\rho+p)\left(1+\epsilon\sqrt{\frac{\rho_{0}}{2\rho}}\right),
\end{eqnarray}

\section{Gravitational baryogenesis in DGP brane cosmology}

From the equation (\ref{1}), the baryon number density to entropy, $\frac{n_{B}}{s}$, for the GB term can be written as \cite{Davoudiasl},
\begin{equation}
\label{4}
\frac{n_{B}}{s}\simeq -\frac{15g_{b}}{4\pi^2g_{*}}\frac{\dot{R}}{M_{*}^2T}\Big{|}_{T_{D}}\, ,
\end{equation}
Here $T_{D}$ is the decoupling temperature. Thus, the derivative of the Ricci scalar play a crucial role in the calculation of the ration of $\frac{n_{B}}{s}$ in the context of GB. In the continuation we must assume that the universe is filled by a perfect fluid with pressure $p$ and energy density $\rho$, which are related as $p=w\rho$. In the GR context, the Ricci scalar can be easily obtained by using the Einstein equations, as follows
\begin{equation}\label{33}
R(t)=\frac{1}{\mu^{2}} \rho(t) (1-3w)\, ,
\end{equation}
where $\mu^2\sim m_{p}^2$ and $m_p$ is the 4-dimensional Planck mass. The ratio of the baryon number density to entropy is calculated by the derivative of the Ricci scalar $\dot{R}$, which in the Einstein's relativity is
\begin{equation}
\label{44}
\dot{R}(t)=\frac{1}{\mu^{2}} \dot{\rho}(t) (1-3w),
\end{equation}
from the above equation it can be seen that in the case of a radiation dominated universe, namely $w=1/3$, the derivative of the Ricci scalar is zero, so the ration of the baryon number density to entropy is zero.

To continue we consider the GB mechanism in  DGP brane scenario by taking $\epsilon=-1$, so by merging equations (\ref{2}) and (\ref{5}), the Ricci scalar $R=12 H^2+6\dot{H}$, can written as
\begin{equation}\label{8}
R(t)=\frac{3 \sqrt{2\rho _0} \sqrt{\rho (t)} (w+1)+(2-6 w) \rho (t)-8 \sqrt{2\rho _0 \rho (t)}}{2 \mu ^2}\, .
\end{equation}
To consider the differences between the DGP brane cosmology and the standard cosmology by comparing the resulting the ratio of baryon number density to entropy in the DGP model, we must calculate $\dot{R}$ as
\begin{equation}\label{9}
\dot{R}(t)=\frac{\left(3 \sqrt{2\rho _0} \sqrt{\rho (t)} (w+1)+4 (1-3 w) \rho (t)-8 \sqrt{2\rho _0 \rho (t)}\right) \dot{\rho}(t)}{4 \mu ^2 \rho (t)}\, .
\end{equation}
Because of the first and third terms in the DGP Ricci scalar of the equation (\ref{8}), the resulting the ratio baryon number density to entropy even for a radiation dominated universe with $w=1/3$ is not zero, which indicate the extra dimension effects of the theory.

Therefor, in the DGP brane cosmology the ratio of baryon number density to entropy in the context of the GB mechanism is not zero, contrary to the GR one. In the following we consider our results by using DGP brane cosmology and we check that our results can be compatible with the observational constraints.

By using the Friedmann equation (\ref{2}) and taking the continuity equation (\ref{3}), with $p=w\rho$, the first order differential equation for $\rho$ is give by
\begin{equation}\label{19}
-\frac{\dot{\rho}(t)}{3 (w+1) \rho (t)}=\frac{1}{ \mu\sqrt{3} }\left(\rho (t)- \sqrt{2\rho_{0} \rho (t)}\right)^{1/2}.
\end{equation}
From the above equation the energy density as a function of the cosmic time is
\begin{equation}\label{11}
\rho (t)=\frac{128 \rho_{0} \mu ^4}{9 \rho_{0}^2 t^4 (w+1)^4-48 \rho_{0} t^2 \mu ^2 (w+1)^2+64 \mu ^4}\, ,
\end{equation}
where we have normalized the energy density to $\rho(t)=2\rho_0$ at $t=0$ and we have considered the integration constant to be zero. Assuming that the universe inters in to the states of thermal equilibrium (quasi-static thermal equilibrium), the total energy density as a function of the temperature is,
\begin{equation}
\label{12}
\rho=\frac{\pi^{2}}{30}g_{*}\, {T}^4\, .
\end{equation}
We can easily obtain the decoupling times $t_{D1}$ and $t_{D2}$ as a function of the decoupling temperature $T_D$, by merging equations (\ref{11}) and (\ref{12}), and the result are,
\begin{equation}\label{13}
t_{D1}=2 \sqrt{\frac{2}{159 \pi }} \sqrt{\frac{53 \pi  (w+1)^2 \mu ^2 \rho _0 g_{*} T_D^4-\sqrt{1590} \sqrt{(w+1)^4 \mu ^4 T_D^4 g_{*}\rho _0^3}}{(w+1)^4 T_D^4 g_{*}\rho _0^2}}\, ,
\end{equation}
\begin{equation}\label{14}
t_{D2}=2 \sqrt{\frac{2}{159 \pi }} \sqrt{\frac{53 \pi  (w+1)^2 \mu ^2 \rho _0 g_{*} T_D^4+\sqrt{1590} \sqrt{(w+1)^4 \mu ^4 T_D^4 g_{*}\rho _0^3}}{(w+1)^4 T_D^4 g_{*}\rho _0^2}}\, .
\end{equation}

From the above equations it can be seen that we have two decoupling time. Thus, from equation (\ref{13}) we can see that the parameter $\rho_0$ is constrained to satisfy the inequality $60\rho_0<\pi ^2 g_* T_D^4$, so it must be at least four orders smaller in comparison to the decoupling temperature. By using equation (\ref{11}) and replacing in equation (\ref{9}), we can write the derivative of $\dot{R}$ as a function of the cosmic time as follows
\begin{equation}\label{77}
\dot{R}=\frac{48 t (w+1)^2 \mu ^2 \rho _0^{7/2} \left(3 (w+1) \sqrt{\frac{\mu ^4 \rho _0}{\left(8 \mu ^2-3 t^2 (w+1)^2 \rho _0\right){}^2}} \sqrt{\frac{\mu ^4 \rho
   _0^2}{\left(8 \mu ^2-3 t^2 (w+1)^2 \rho _0\right){}^2}}-\frac{8 \mu ^4 \sqrt{\rho _0} \left(4 \sqrt{\frac{\mu ^4 \rho _0^2}{\left(8 \mu ^2-3 t^2 (w+1)^2 \rho
   _0\right){}^2}} (3 w-1)+\rho _0\right)}{\left(8 \mu ^2-3 t^2 (w+1)^2 \rho _0\right){}^2}\right)}{\left(\frac{\mu ^4 \rho _0^2}{\left(8 \mu ^2-3 t^2 (w+1)^2 \rho
   _0\right){}^2}\right){}^{3/2} \left(8 \mu ^2-3 t^2 (w+1)^2 \rho _0\right){}^3}\, ,
\end{equation}
so by using the first decoupling time $t_{D1}$ (\ref{13}), the term $\dot{R}$ as a function of the decoupling temperature is,
\begin{eqnarray}\label{55}
\dot{R}&=&-\frac{\pi ^{3/2} \sqrt{g_\ast T_{D}^4 (w+1)^4 \mu ^4 \rho _0^3} \left(-2 \sqrt{5} g_\ast \pi  (3 w-1) T_{D}^4-15 \sqrt{3} \sqrt{g_\ast T_{D}^4} (w+1) \sqrt{\rho _0}+40 \sqrt{3} \sqrt{g_\ast T_{D}^4 \rho_0}\right) }{300 \sqrt{2} \mu ^6 \rho _0}\\&&\times\nonumber
\sqrt{\frac{\pi  (w+1)^2 \rho _0 \mu ^2-\frac{2 \sqrt{15} \sqrt{g_\ast T_{D}^4 (w+1)^4 \mu ^4 \rho _0^3}}{g_\ast T_{D}^4}}{(w+1)^4 \rho _0^2}}\, .
\end{eqnarray}
 By replacing $\dot{R}$ from equation (\ref{55}) in equation (\ref{4}), the final expression of baryon number density to entropy in the DGP brane model is
\begin{eqnarray}\label{88}
\frac{n_{B}}{s}&\simeq&\frac{g_b \sqrt{(w+1)^4 \mu ^4 T_D^4 \rho _0^3 g_*} \left(-2 \sqrt{5} \pi  (3 w-1) g_* T_D^4-15 \sqrt{3} (w+1) \sqrt{\rho _0} \sqrt{T_D^4 g_*}+40 \sqrt{3}\sqrt{T_D^4 \rho _0 g_*}\right)}{1200 \sqrt{2 \pi } \mu ^6 T_D \rho _0 g_* M_*^2}\\&&\times\nonumber
\sqrt{\frac{\pi  (w+1)^2 \rho _0 \mu ^2-\frac{2 \sqrt{15} \sqrt{g_\ast T_{D}^4 (w+1)^4 \mu ^4 \rho _0^3}}{g_\ast T_{D}^4}}{(w+1)^4 \rho _0^2}}\, .
\end{eqnarray}
Here we discuss under which  situations the resulting ratio of baryon number density to entropy can be compatible with the theoretical bound $n_{B}/s\preceq 9 \times 10^{-11}$. We use Planck units for simplicity, and we choose the cutoff scale $M_{*}$ is equal to $M_{*}=10^{12}$GeV, and also that the decoupling temperature is $T_{D}=M_{I}=2\times 10^{16}$ GeV, where $M_{I}$ is the upper
bound for tensor-mode fluctuations constraints on the inflationary scale.
Also we set that $g_{b}\simeq {\cal O}(1)$ and also that $g_{*}=106$ which is the total number of the effectively massless particles in the early universe. Finally, we assume that $\rho_{0}\simeq 1.56\times10^{49}$GeV$^4$ and we can use the effective equation of state parameter $w$, to obtain the ratio of baryon number density to entropy. For $w=1/3$, the radiation dominated universe, in this case ratio of baryon number density to entropy obtained by gravitational baryogenesis (\ref{88}) approximately is equal to $n_B/s\simeq 8.38193\times 10^{-11}$, which is compatible with the observational bounds. Also by choosing $w=0$, which corresponds to the matter dominated epoch, the ratio of baryon number density to entropy is $n_B/s\simeq -0.00123978$, which is not compatible with the observational bounds.

By repeating the above calculations for GB with the second decoupling time $t_{D2}$ for $w=1/3$ we obtain the ratio of baryon number density to entropy obtained by GB mechanism is $n_B/s\simeq 9.53281\times 10^{-11}$, which is compatible with the observational bounds. Also for matter dominated $w=0$ universe, the resulting ratio of baryon number density to entropy is $n_B/s\simeq 5.66208\times 10^{-11}$, which is again compatible with the observational bounds.
Note that in the DGP brane cosmology the ratio of baryon number density to entropy extremely depends on the parameter $\rho_0$, which for consistency has to be roughly smaller than the fourth power of the decoupling temperature, that is $\rho_{0}\prec T_{D}^4$.

\section{Conclusions}

In this paper we have studied the effects of brane cosmology in context of DGP model via GB mechanism. Also, we have explicitly obtained the baryon number density to entropy ratio for DGP brane in background of FLRW universe, to do this the derivative of Ricci scalar, $\dot{R}$, is calculated. The crucial point of our work is the fact that even in the case of a radiation dominated epoch, the ratio of baryon number density to entropy is not zero, which is in contrast to the standard cosmology. Thus, if the GB mechanism is discussed a viable baryon asymmetry generating mechanism, the DGP brane cosmology extremely affect the amount of imbalance of matter and antimatter in the early universe. We have considered the cosmological evolutions in the context DGP brane model that is consistent with the observational bounds on gravitational baryogenesis. As we have shown, by fixing the parameters of the model the observational bounds on the ratio of baryon number density to entropy can be achieved. Finally, from the expression of the decoupling time, we have obtained the critical density parameter must be smaller than the fourth power of the decoupling temperature, namely $\rho_{0}\prec T_{D}^4$.



\begin{thebibliography}{99}


\bibitem{devali}
D. Dvali, G. Gabadadze and M. Porrati, Phys. Lett. B \textbf{485} (2000) 208 ;\\ C. Charmousis,
R. Gregory, N. Kaloper and A. Padilla, JHEP \textbf{0610 }(2006) 066;\\ R. Gregory, N. Kaloper, R.
C. Myers, A. Padilla, JHEP \textbf{0710} (2007) 069;\\ D. Gorbunov, K. Koyama and S. Sibiryakov,
Phys. Rev. D \textbf{73} (2006) 044016.


\bibitem{Deffayet} C. Deffayet, Phys. Lett. B  \textbf{199} (2001) 502.

\bibitem{Lue} A. Lue and G. D. Starkman, Phys. Rev. D \textbf{70} (2004) 101501(R).

\bibitem{26} L. P. Chimento, R. Lazkoz, R. Maartens and I. Quiros, JCAP \textbf{0609}
(2006) 004.

\bibitem{30} M. Bouhmadi-Lopez, R. Maartens and D. Wands, Phys. Rev. D\textbf{ 70} (2004) 123519.
\bibitem{31} R. Cai and H. Zhang, JCAP \textbf{0408} (2004) 017.
\bibitem{32} E. Papantonopoulos and V. Zamarias, JCAP \textbf{0410} (2004)  001.
\bibitem{33} H. Zhang and R. Cai, JCAP \textbf{0408} (2004) 017.
\bibitem{34} H. Zhang and Z. Zhu, Phys. Lett. B \textbf{641} (2006) 405 .
\bibitem{35} S. del Campo and R. Herrera, Phys. Lett. B \textbf{653 }(2007) 122.


\bibitem{Davoudiasl}
  H.~Davoudiasl, R.~Kitano, G.~D.~Kribs, H.~Murayama and P.~J.~Steinhardt,
    Phys.\ Rev.\ Lett.\  {\bf 93} (2004) 201301.

\bibitem{Bennett}
  C.~L.~Bennett {\it et al.} [WMAP Collaboration],
    Astrophys.\ J.\ Suppl.\  {\bf 148} (2003) 1.


\bibitem{Burles}
  S.~Burles, K.~M.~Nollett and M.~S.~Turner,

  Phys.\ Rev.\ D {\bf 63} (2001) 063512.

\bibitem{sakhar}
A.~D.~Sakharov,
  Pisma Zh.\ Eksp.\ Teor.\ Fiz.\  {\bf 5}, 32 (1967)
  [JETP Lett.\  {\bf 5}(1967) 24 ]
  [Sov.\ Phys.\ Usp.\  {\bf 34} (1991) 392]
  [Usp.\ Fiz.\ Nauk {\bf 161} (1991) 61].


\bibitem{Lambiase}
  G.~Lambiase and G.~Scarpetta,   Phys.\ Rev.\ D {\bf 74} (2006) 087504.

\bibitem{Ramos}M. P. L. P. Ramos, J. P$\acute{a}$ramos, Phys. Rev. D \textbf{96}  (2017) 104024.
\bibitem{Koyama}T. Shiromizu, K. Koyama, JCAP \textbf{0407} (2004) 011.
\bibitem{Odintsov} S. D. Odintsov and V. K. Oikonomou, EPL \textbf{116} (2016) 49001, arXiv:1610.02533.
\bibitem{sari} V.~K.~Oikonomou and E.~N.~Saridakis, Phys. Rev. D \textbf{94} (2016) 124005 .
 \bibitem{Bento}M. C. Bento, R. Gonzalez Felipe and N. M. C. Santos, Phys. Rev. D \textbf{71} (2005) 123517.
\bibitem{Odintsov2}S. D. Odintsov and V. K. Oikonomou, Phys. Lett. B \textbf{760} (2016) 259.
\bibitem{Oikonomou}V. K. Oikonomou, Supriya Pan and Rafael C. Nunes, arXiv:1610.01453.
\bibitem{Rudra}S. Maity and P. Rudra, arXiv:1802.00313.
\bibitem{1} G.~Lambiase, S.~Mohanty and A.~R.~Prasanna, Int.\ J.\ Mod.\ Phys.\ D {\bf 22} (2013) 1330030.

\bibitem{2}
  G.~Lambiase, Phys.\ Lett.\ B {\bf 642} (2006) 9.

\bibitem{3}
  H.~Li, M.~z.~Li and X.~m.~Zhang, Phys.\ Rev.\ D {\bf 70} (2004) 047302.

\bibitem{4}
  L.~Pizza,  arXiv:1506.08321 [gr-qc].

  \bibitem{5}J.~A.~S.~Lima and D.~Singleton, arXiv:1610.01591 [gr-qc].


\bibitem{6} V.~K.~Oikonomou, Int.\ J.\ Geom.\ Meth.\ Mod.\ Phys.\  {\bf 13} (2016) 1650033.

\bibitem{Maeda}K. Maeda, S. Mizuno and T. Torii, Phys. Rev. D \textbf{68} (2003) 024033.
\bibitem{wu}K. Zhang, P. Wu and H. Yu, Phys. Lett. B \textbf{690} (2010) 229.



\end{thebibliography}
\end{document}